# General Theories and Features for Interfacial Thermal Transport

Hangbo Zhou (周杭波), Gang Zhang (张刚)[1]

Institute of High Performance Computing, A*STAR, Singapore 138632

## Abstract

A clear understanding and proper control of interfacial thermal transport is important in nanoscale device. In this review, we first discuss the theoretical methods to handle the interfacial thermal transport problem, such as the macroscopic model, molecular dynamics, lattice dynamics and modern quantum transport theories. Then we discuss various effects that can significantly affect the interfacial thermal transport, such as the formation of chemical bonds at interface, defects and interface roughness, strain and substrates, atomic species and mass ratios, structural orientations. Then importantly, we analyze the role of inelastic scatterings at the interface, and discuss its application in thermal rectifications. Finally, the challenges and promising directions are discussed.

---

[1] Email: zhangg@ihpc.a-star.edu.sg

**Introduction**

The continuous scale down of nanotechnology has simultaneously increased the density of interfaces in nano-devices. The interface properties, which can be very different from its bulk counterpart, can be utilized to improve the material performance. For example, interface engineering has merged for the tailoring of intrinsic electron properties in monolayer molybdenum disulfide ($MoS_2$)[1, 2]. Therefore, a clear understanding of interface properties is highly demanded for the discovery and design of high-performance nano-devices.

Thermal transport in nanostructures has continuously attracted intensive research interests[3-9]. The thermal transport at the interface is of particular importance because the huge density of heat generation requires efficient ways of heat dissipation[10], but the interface creates extra barrier to the heat flow[11, 12]. For example, the thermal resistance of the interface between carbon nanotube and its electrodes is a critical issue for carbon nanotube based applications[13]. This interfacial thermal resistance, also called Kapitzal resistance or thermal boundary resistance, was first found between helium and solid in 1941[14]. Since then, theories of interfacial thermal transport have been restlessly developing, from the acoustic mismatch match model, diffusive mismatch model, molecular dynamics to modern quantum simulations. Now it has been realized that the interfacial thermal resistance is caused by various reasons at difference conditions, such as mismatch of atomic vibrational energy, localization of heat carriers at interfaces and extra scatterings experienced by heat carriers, to list a few. Hence the investigation of interfacial thermal resistance becomes a comprehensive problem, depending on different situations.

Despite its negative impacts on heat dissipation, interfacial thermal resistance has also been revealed useful in many recent studies. Firstly, the interfaces provide additional methods to modulate the heat flow, and hence it can be used for heat management. Many factors, such as lattice/mass disorders, formation of chemical bonds, interface roughness and defects can substantially affects the interfacial thermal resistance [15]. For example, the presence of interface modulation enables the achievement of graphene-based thermal modulators [16]. Some interfaces like graphene-$MoS_2$ in-plane contacts, the interfacial thermal resistance can be modulated by introducing vacancies, thereby the magnitude of heat current is adjustable by varying vacancy concentrations [17]. A recent study reveals that the interfacial thermal resistance between metal and dielectric materials can be tuned through insertion of appreciate interlayer materials with strong electron-phonon interaction [18]. All these studies show that interfaces give more degrees of freedom to control the heat flow.

Another important application of interfacial thermal resistance is in designs of thermoelectric materials [19]. High-performance thermoelectric materials require high electric conductivity but low thermal conductivity in order to achieve high heat-to-electric conversion efficiency. Through creation of boundaries, it is possible to significantly suppress the thermal conductivity but make the electronic properties unchanged much. These interfaces can be realized through nano-grains [20] and supperlattice with isotopes [21]. Simulation shows that it gives an efficiency way to enhance the thermoelectric performance.

Interfaces also provide opportunities to create thermal rectifiers. Thermal rectification is the asymmetry of heat flow between forward and backward directions under same temperature bias. It is the underlying mechanism to realize thermal diode [22] and phononics [23]. Thermal rectification can be achieved through interfaces between dissimilar lattices [24].

In this review, we do not aim to cover the extensive literatures on interfacial thermal transport, but to capture some advances from the recent studies, including the role of interfacial atomic details, chemical bonds, strains, defects, and inelastic scatterings in interfacial thermal transport. We also discuss the interface-induced thermal rectifiers. In section 2 we will discuss the simulation methods for interfacial thermal transport. In section 3 we will discuss various effects on interfacial thermal transport, including the interfacial bonds, strains, defects and so on. In section 4, effects of inelastic scatterings and its applications in thermal rectifications are addressed. In section 5 we will summarize and give perspectives.

## 2. Theory and Simulation methods
### 2.1 Macroscopic theories

After the discovery of interfacial thermal resistance, several macroscopic theories have been proposed to explain underlying reasons of the interfacial thermal resistance. One of them is the acoustic mismatch model (AMM) [25]. The idea of AMM is that the interfacial thermal resistance is due to the mismatch of acoustic impedances between two materials. This mismatch is due to the different acoustic propagation properties of the two bulk materials that form the interface, such as their different sound speeds. In the AMM model, the details of interface, such as geometry, orientation and chemical bonds, are not taken into consideration. Interfacial thermal resistance is completely estimated from the acoustic properties of the bulk materials themselves. AMM gives the transmission coefficient for an acoustic mode that transmit through the interface formed by A and B. For a mode that is normal to the interface, the transmission coefficient is given by $t_{AB} = 4Z_A Z_B/(Z_A + Z_B)^2$, where $Z_A$ and $Z_B$ are the acoustic impedance of material A and B respectively. From this expression we can immediately find that the transmission is symmetry between A and B. In other words, interfacial thermal conductance is predicted to be the same if we reverse the direction of heat flux. We will see in Section 4 that this is not necessarily true and thermal rectification is possible due to interfaces. Despite its simplicity, AMM do explain the interfacial thermal conductance well at the low temperature [14]. Its prediction that the interfacial thermal conductance is proportional to the cube of temperature in the low temperature limit is in good agreement with early experiments.

Because AMM assumes that the phonon modes do not experience any scatterings at the interface, it normally underestimates the interfacial thermal resistance. As a complementary theory, diffusion mismatch theory (DMM) is proposed such that it assumes all the phonons at the interfaces are completely scattered. It also estimates transmission probabilities of phonons at the interface. Phonons loss all the memories about their previous states. As a result, the transmission probability is not related to the incident angle, the group velocity or the wave front of the phonon modes. The phonons arriving at the interface have chances to be scattered into phonon states at the both sides of the interface. As a result, the transmission phonon energy is linearly dependent only on the density of states. The interfacial thermal resistance predicted by DMM reaches minimum when the overlap of the density of states is maximized. Due to the assumption of complete scatterings, DMM usually overestimate the interfacial thermal resistance, and it is normally works better in high temperature regimes.

Both AMM and DMM ignore the atomic details and structures of the interface itself, by considering only the vibration properties of two bulk materials that forms the interface. So they can only provide a qualitative estimation. Atomic-level theories or simulation tools are required for more accurate calculation. Furthermore, DMM only considers elastic scatterings, which means that the energy of each phonon across the interface does not change. Actually the inelastic scattering at the interface can become important when the mismatch of phonon spectra is high, and it can cause interesting phenomenon such as thermal rectification. We will address these issues in Section 4.

Improved models based on AMM and DMM has been proposed to eliminate the crucial assumptions or to consider other factors that AMM and DMM are not able to do. For example, both AMM and DMM assume linear dispersion relation. A modified DMM has been proposed, which is able to consider the full dispersion relation in lattice, widely broadening the application regime of DMM [26]. Other improvements are also proposed such as to consider electron-phonon scatterings, disorders and other phonon scatterings [27].

Besides the AMM and DMM, there are also other macroscopic theories to describe interfacial thermal transport. For example, analytical expressions of heat transport across flat interface, based on the surface displacement, is presented in Ref.[28]. Beyond the transport properties of phonons, interfacial thermal transport due to electron-phonon interactions is also presented [29], which is important in metal-nonmetal interfaces. Overall, due to its capability of capture the general picture, macroscopic theories are still developing.

## 2.2 Molecular dynamic simulation

Though the macroscopic theories provide the general picture of the interfacial thermal transport, it normally oversimplifies the complexity of the interface. The atomic details of the interface, especially in nanoscale structures, can significantly influence the thermal resistance. Forms of chemical bonds, defects and atomic species at interfaces are important factors. Molecular dynamics, an atomic-level simulation, is a powerful tool to analyze these effects.

The most commonly used method of molecular dynamics to simulate interfacial thermal transport is the non-equilibrium molecular dynamics (NEMD), or the so-called direct method. In this method, a temperature bias is directly applied to the materials that forms the interface, and then system evolves according to the inter-atomic potentials. Normally the evolution time is about few nanoseconds to microsecond, depending on the size of the system and the required accuracy. During the evolution, the total energy that transferred from one side of interface to the other side is counted. Therefore, the heat current, J, is known. Interfacial thermal conductance (ITC), σ, can be evaluated as

$$\sigma = \frac{J}{\Delta T}$$

where $\Delta T$ is the temperature difference across the interface. We need to emphasize that the temperature difference at the interface is not the applied temperature difference because the materials themselves has thermal resistance. Thereby temperature gradually decrease along the materials, then a sudden drop of temperature occurs at the interface, and it slowly decrease again along the other side. The amount of temperature that drops at the interface is the $\Delta T$ in the formula. If interface does not exist, then $\Delta T = 0$, it recovers the fact that interfacial thermal conductance is infinite, in other words, the interfacial thermal resistance is zero.

The principle of NEMD is straightforward. However, an important question is what the temperature is. A clear answer to this question is not easy due to its non-equilibrium nature. The applied temperature is often characterized by controlling the distribution of velocities. It can be realized by connecting to a heat bath. Heat baths are in thermal equilibrium so that temperature is well defined. Sophisticated heat bath, such white-noise bath, can be implemented. The ambiguous part is how to measure the local temperature along the material and the interface, because they are in non-equilibrium state. The simplest way is to use the equal partition formula that the temperature is proportional to the average of kinetic energy $\langle mv^2 \rangle = 3k_B T$, by assuming that the system is large enough and local equilibrium exists. Discussion of other methods, such as fitting the velocity profile to temperature-related distributions, using temperature probes, can be found in the literature [11].

MD envelops all the information in a single quantity, the interfacial thermal conductance. To analyze transport mechanism, one needs to investigate in detail. For example, MD allows calculation of local thermal current to see the efficiency of each transport channels. Modal analysis is also proposed to gauge the effects of inelastic scattering [30].

NEMD provides an atomic-level simulation and it is widely used for the simulation of interfacial thermal transport. It can fully consider detailed factors of the interface such as defects, strains, and chemical bonding. The accuracy of NEMD is normally determined by the accuracy of the inter-atomic potential and the simulation time. It can also be used as benchmark for the other theoretical predictions [27]. The limitation of NEMD is that it is entirely classical. Specially the meaning of "classical" comes in two parts. The first one is that the evaluation is completely follows Newtonian dynamics. Hence all the phonons are equally excited, the allowed amplitude of atomic vibration is continuous. Hence the transport quanta, which has been observed experimentally, is impossible to be predicted. The second is the phonon distribution is classical. The phonons do not obey Bose-Einstein distribution even at equilibrium. Therefore, NEMD is expected to breakdown at low temperature. At the regime when the quantum effects become important, quantum theories are necessary to handle to the transport properties.

**2.3 Lattice dynamics**

For interfaces formed by crystals, lattice dynamics is an alternative method to evaluate the interfacial thermal conductance [27, 31-33]. The key quantity of lattice dynamics is the displacement of atom from its equilibrium position. The basic idea is to solve the equation of motion of these displacements. The equation of motion is governed by the Hamiltonian, or the potential energy of the atomic displacements. In harmonic approximation, the potential energy is simplified to be quadratic, characterized by spring constant matrix. In periodic lattice, the displacement of the atoms can be decomposed to independent vibration modes, after taking Fourier transformation. This is often called as normal modes decomposition. The advantage of the decomposition is that each mode becomes independent of each other, in other words, they are decoupled. The analysis of atomic displacement is equivalent to the analysis of the phonon wave-vector, often denoted as $q$. Since each mode are decoupled, every phonon mode $q$ is associated with a specific frequency $\omega$. The relationship between $\omega$ and $q$ is called as phonon dispersion relation. We need to remind that the concept of phonon dispersion relation is under the assumption of harmonic approximation. In anharmonic case modes are not independent to each other. In simple words the vibration is not time periodic and then frequency is not well-defined (The wave vector is still well-defined due to existence of lattice periodicity). In lattice, there are usually more than one degree of freedom in a unit cell. Then each degree of freedom will contribute to a phonon branch in the dispersion relation. It means that at each wave-vector $q$, there exists $n$ vibrational patterns in a unit cell, where $n$ is the number of degrees of freedom. Those modes that all the atomics inside the unit cell vibrates coherently are called acoustic modes. If there is relative movement of atomic vibration inside a unit cell, then they are optical modes. The three common acoustic modes are longitude acoustic (LA) mode and transverse acoustic (TA) mode. They correspond to the coherent vibration that parallel and perpendicular to the wave-vector $q$, respectively. For two-dimensional materials, one of the transverse modes with out-of-plane vibration is often called ZA mode, or flexural mode.

The concepts of lattice dynamics can be employed to analyze the transmission properties of phonon mode at the interface. Such theories are called scattering boundary theories. The amount of transmitted energy can be evaluated by solving the equation of motion of the atoms at the boundary. Scattering boundary is useful to reveal the transport mechanisms through modal analysis. To characterize the transmission properties of interface, wave-package method becomes useful [34]. In this method, a wave package of phonons modes with certain frequency is generated at one side of interface, and then it propagates to the other side. One can then analyze the transmitted and reflected energy of this wave package, in order to find the transport properties of such specific mode.

Wave packet method provides direct information about phonon transmission. The snapshots of wave packets of four representative acoustic phonon modes are shown in Figure 1. It is clear that the wave packets from LA and TA bands nearly experience no reflection and no variation in group velocity. In contrast, the wave packet from ZA mode with frequency 2.5 THz is totally reflected by the interface. Another ZA packet with frequency of 9.0 THz can transmit across the interface, but there is a clear reflected wave packet. This is consistent with the analysis of the phonon dispersion. As is shown in Figure 2, the in-plane phonon bands in encased graphene fit well with those in suspended graphene, but the flexural bands are flattened and shifted in encased graphene. Because the substrates break the translational invariance of graphene at out-of-plane direction, the frequency of ZA mode near Γ point shifts to 6 THz. This shift leads to mismatch of phonon dispersions at two sides. From the viewpoint of elastic interfacial scattering, the in-plane phonons can transmit across the interface without being scattered (Figure 1a, b) due to the perfect match of in-plane phonon dispersions of both graphene sections, whereas the ZA phonons with frequency smaller than 6THz are totally reflected (Figure 1c) by the interface due to the absence of phonon modes at the other side. However, ZA phonons with frequency larger than 6THz are partly reflected (Figure 1d) because of the different group velocities of its counterpart on the other side.

**2.4 Quantum theories**

In the nanoscale or low temperature regime, one needs to implement quantum distributions and evolutions to calculate the interfacial thermal transport properties. The difficulty is that one has to properly introduce the concept of temperature, a thermodynamic quantity, into the quantum dynamics. For that reason, quantum heat bath is necessary where the particles inside follows the Bose-Einstein distribution for Bosons and Fermi-Dirac distribution for Fermions. The thermodynamics requires that the heat bath should have infinite degrees of freedom. It introduces difficulties to quantum mechanical treatment because the infinite limit needs to be properly taken care of.

A well-developed quantum theory to handle thermal transport is the non-equilibrium Green's function (NEGF) formalism [35]. It is based on the quantization of the lattice dynamics and scattering theories. In the elastic regime, it recovers the Landauer formula. The Landauer formula gives a viewpoint that thermal transport is described as the transmission of heat carriers between two equilibrium baths. For phonons, it can be written as

$$I = \int_0^\infty \frac{d\omega}{2\pi} \hbar\omega\, T(\omega)[f_L(\omega) - f_R(\omega)]$$

where $T(\omega)$ is the transmission coefficient, $\hbar\omega$ gives the energy of phonons and $f_{L(R)}(\omega)$ is the Bose-Einstein distribution of the left and right regime respectively. The transmission coefficient can be calculated starting from the scattering boundary theories, or by using the Green's function technique. We need to emphasize that this formula is to evaluate elastic transport, which means the phonons do not loss or gain energy during travel. For inelastic transport, NEGF still gives a formal expression to evaluate the thermal transport [36], which is beyond the scope of Landauer picture.

The NEGF formalism is based on junction setup, where two (or more) heat baths are connected to a non-equilibrium center. The transmission coefficient measures the probability of phonon transmits from one heat bath to another through the center. Therefore, to evaluate the interfacial thermal transport, one needs to extend the interface and then regard it as a scattering center [37, 38]. Normally it will increase the computational complexity. The formalism that directly calculates the interfacial thermal transport, where the heat baths are connected directly, has been developed [39, 40] and it is promising to apply it to real materials.

Technological improvements based of NEGF formalism is continuously developing, For example, the implement of inelastic scatterings, such as electron-phonon scattering [29, 41], phonon-phonon scattering [42], has been developed. The modal analysis under the NEGF framework has also been formulated [43]. Integration of NEGF and MD has also been proposed to handle complex structure of interface, where MD is used to simulate the atomic reconstruction and relaxation [44]. Beyond NEGF, other quantum theories based on wave function picture, has been developed to handle interface thermal transport [45]. It is proposed that the interfacial heat flux can be evaluated from the displacement fluctuations of the atoms at the interface [46].

## 3. Factors that have impacts on ITC.
### 3.1 Interfacial bonds

The strength of the bonding at the interface can directly affect the interfacial thermal conductance. In some interfaces, the bonding of the interface is very week. For example, they are only connected through van der Waals interactions. AMM, also NEGF, predicts that in the weak bonding strength limit, ITC is proportional to the square of bonding strength [38]. However, in the moderate bonding strength regime, ITC can increase linearly with bonding strength. Furthermore, if the bonding is too strong, even stronger than the bonding inside the bulk material, then ITC may be suppressed with increase of bonding strength.

Many NEMD studies imply that the formations of chemical bonds have significant influence on the ITC. The ITC between $MoS_2$ and electrodes is studied [47], where covalent bond can be formed. It was found that the covalent bonds serve as thermal channels for phonon to transmit over the interface. ITC becomes proportional to the density of bonds at the interface. Another study of ITC between silicon and a vertical carbon nanotube shows that the ITC can increase two order of magnitude if the chemical bonds are formed at the interface [48]. Similarly, a study on $MoS_2$-graphene in-plane interface was investigated by using NEMD. As shown in Figure 3, at the interface, covalent Mo-C bonds are formed, whose strength is comparable to the graphene-metal bonds [9]. Such strong Mo-C bonds also provide channels for phonon to transport. The ITC of the graphene-MoS2 interface is around $2.5 \times 10^8$ $WK^{-1}m^{-2}$ at room temperature, which is comparable with that of chemically-bonded graphene-metal interfaces, indicating that $MoS_2$-graphene in-plane heterostructure may provide a viable solution for thermal management in $MoS_2$-based electronic devices. Compared with the pristine interface, there is about 12% reduction of ITC at interface with point defects.

The ITC can be controlled through the modulation of the density of interfacial chemical bonds. A study on silicon oxide-silicon ($SiO_2$-Si) interface reveals that interfacial bond strength can affect its sensitivity to the atomic structures [49]. For weak bonds ITC is very sensitive to the atomic structure of the interface, while in strong bonds regime, it becomes insensitive to the detailed structures. Other systems, such as interface between suspended and encased graphene [50], self-assembled SAM-SI interface [51] and metal-insulator interface, also shows strong bond-strength dependence of ITC. In addition to the bond-strength, an NEGF study of graphene/hexagonal boron nitride (h-BN) interface shows that the type of bonds can also affects the thermal transmission [43]. The carbon-nitrogen bonds in the zigzag direction transmit high frequency phonons more efficiently than the carbon-boron bonds.

Atomic defects, such as atomic vacancies, are always present in materials. The vacancies disrupt lattice structures and cause additional scattering to phonon transport. Similar, atomic vacancies at the interface will also affect the ITC remarkably. As shown in Figure 4, the total ITC decreases linearly with number of vacancies at the contact. Because the introduction of atomic vacancies at the interface corresponds to the decrease in the density of covalent bonds, thus the ITC increases linearly with the number of bonds at the interface. This linear dependence indicates that each chemical bond serves as an independent heat transport channel with a constant thermal conductance.

**3.2 Defects and surface roughness**

Interfaces are the places that are easy to introduce defects, due to the mismatch of the lattice, the release of the stress and the contamination during the formation of the interfaces. Then one would ask what the effect of defects on the ITC is. Usually defects will introduce imperfection to the lattice structure, increase the surface roughness, cause additional scatterings to the phonons and then reduce the ITC. While high frequency phonons can be scattered by the lattice mismatch, the defects are able to scatter low frequency phonons and hence reduce the ITC [44]. However, a recently study on the graphene/h-BN interface found an abnormal behavior that defects are able to enhance the ITC [52]. As shown in Figure 5, for defect-free Gr/*h*-BN heterostructure, the local heat flux in the central sections is almost uniformly distributed, and local heat flux at the edge is about 25% lower than that at the center as consequence of edge scattering-induced localized phonon modes. However, for the BN-C$_{5|7}$ heterostructure with interface defect, although the local heat flux at the defect positions reduces with respect to the perfect interface, the local heat flux density at most part of the Gr/*h*-BN boundary increases upon introducing 5|7 defects. As the heat flux reduction at the local position of defects is offset by the large increase in heat flux at positions without defects, there is an overall enhancement in total ITC.

Interface roughness is another important factor influencing the ITC. Experimentally the interface roughness can be controlled by growth condition and formation process. Generally the interface roughness has similar effects as defects, which cause scatterings and localization of phonons, especially for high-frequency phonons [15, 53]. However, the phenomenon that enhancement of phonon transmission by surface roughness has been reported at Silicon-Germanium (Si-Ge) interface [54]. As shown in Figure 6, a significant increase in interfacial thermal conductance is observed in interface with 6-layer Gaussian roughness, as a consequence of the increase in phonon transmission. The roughness can soften the abrupt change of acoustic impedance, resulting in enhancement of phonon transmission.

### 3.3 External forces

We first look at the effects of applied strain to ITC. Stain is a commonly used way to modulate the interatomic force constants and hence change the thermal transport properties[43, 55-57]. Using graphene junctions with different mass distributions as an example, Pei et al. studied the effect of strain on interfacial thermal resistance. Strong phonon scattering at the interface results in sharp temperature jump in the temperature profile. As shown in Figure 7, tensile strain is able to decrease the ITC. This phenomenon can be understood from the phonon power spectra. From phonon density of states spectra near the interface as shown in Figure 7, it is clear that the low frequency peak (at ~16 THz) is insensitive to strain, while the high frequency peaks red shift with tensile strain. For atoms on the two sides of interface, the shift rate is different, and reduces the overlap between phonon spectra. This is the underlying mechanism for the increase of interfacial thermal resistance. Similar, in graphene/MoS2 interface, on the opposite, a tensile strain will reduce the ITC while a compressive strain can increase the ITC. On the other side, if strain shifts the phonon bands, and such shift causes better alignment of phonon spectra, it will general improve the ITC [20, 56]. The concept of overlapping of phonon density of state is a simple way to evaluate ITC [13, 56]. A recent study shows that, to be more accurate, the overlap of vibration power spectrum of the contact atoms should be considered [58]. It is important to emphasize that the idea of evaluate phonon transmission through the analysis of spectra overlap is restricted to elastic scattering, and within the weak interfacial coupling regime. Inelastic scatterings can change the frequency of phonons at the interfaces; thereby phonon transmission is even possible without the alignment of phonon bands. Actually inelastic scattering can even cause thermal rectification, breaking the symmetry between the forward and backward heat flux from spectra overlapping prediction.

Another kind of external forces is the van der Waals interaction, through the substrates or other coating materials for two-dimensional materials. These interactions provide onsite potentials to the atoms in the material and hence greatly affect their out-of-plane vibration. Taking graphene as an example, the flexural modes (ZA modes) is the major heat carrier in suspended graphene and it is significantly suppressed once it is encased. Interestingly, even for one graphene sheet, if half section is supported and the other half is suspended, there still exists interfacial thermal resistance inside graphene sheet [50]. The system structure is shown in Figure 8, a single layer graphene half encased by another two graphene layer structures, which serve as external perturbation. A large temperature jump $\delta T$ is observed at the interface, indicating the existence of interfacial thermal resistance. The interfacial thermal resistance is $\sim 5.37 \times 10^{-11} m^2 K/W$, which is comparable with interfacial thermal resistance at tilt grain boundaries in graphene, and provides the first demonstration that due to the interfacial thermal resistance originating from inhomogeneous external potential is of remarkable influence on thermal transport, in the absence of any contact roughness or misorientation.

The phenomenon demonstrated in the encased and suspended graphene junction closely relates to graphene's two-dimensional topology, so that the weak graphene-substrate interaction can directly influence its thermo-mechanical property. It is worth emphasizing that using this effect, graphene-based thermal modulators has been proposed. The clamp-graphene distance, controlled by external pressure, is able to module the thermal transport of graphene [16]. In boron nanoribbons, it was found that the van der Waals interactions cause the enhancement of thermal conduction [59]. It is suggested that inelastic scatterings take the responsibility.

**3.4 Atomic species and mass ratio**

The atomic species of the interface can significantly affects the ITC. For example, the graphene-MoS2 has much lower ITC than graphene/graphene or MoS2/MoS2 interface [57]. It is also reported that the carbon atoms in graphene/h-BN interface reduces the ITC [60]. In other cases, even the species of atoms are the same, the ITC can be affected by introducing isotope atoms [56]. The difference of atomic mass cause phonon scatterings and then reduce the thermal transport. Hence it has been proposed that interfacial thermal transport between two dissimilar materials can be tuned by modulating their relative mass ratio [61]. As the atomic species at the interface can efficiently affect the ITC, an interesting idea is to use molecular cross-linker to modulate thermal transport. It has been found that a cross-linker between two graphene nanoribbon can effective transmit out-of-plane modes and it filters the in-plane modes [62].

### 3.5 Structural orientations and incident angles

ITC is often closely related to the lattice orientations, even when the materials themselves are isotropic. In graphene/h-BN interface, it is found that zigzag interfaces cause stronger reduction of ITC, due to enhanced phonon localization. The interfacial structure can also affect the orientation of transmitted phonons [43]. Besides the alignment of interfacial structure, the incident angle of phonons is also important to determine whether it is transmitted or reflected. A simulation of Si/Ge interface shows that the phonon transmission changes smoothly by varying the incident angle, and a critical angle exists [63].

### 3.6 Size effects

Due to contribution from acoustic modes that possess long wavelength, size will also affects the interfacial thermal conductance [64]. In the study of interfaces formed by suspended and encased graphene, it is found that ITC will increase with the increase of the system length, and then gradually saturates [50]. Similar results are observed from the interfaces formed due to grain boundaries of graphene; where a decrease in size will significant suppress the ITC [65]. It has also been suggested that the size effects is more significant in weakly-coupled interfaces, but less important in strongly coupled interfaces [49].

**3.7 The role of inelastic scatterings**

Inelastic scattering means that the energy of phonon changes during the scattering. For example, in the three-phonon process an incident phonon can split into two phonons, or two phonons can be combined into a single one. If inelastic scattering happens, then the energy can transfer from the low frequency phonons at one side of interface to high frequency phonons at the other side [66].

Normally inelastic scatterings suppress the ITC because it introduces additional scatterings to the phonons, and hence reduce the mean free path of the phonons. Inelastic scatterings thermalize the phonons and cause them to obey the Bose-Einstein distribution in equilibrium condition. However, experiments on metal-diamond interfaces show that the ITC is extremely high[67, 68], 100 times larger than the calculation of lattice dynamics. It suggests that inelastic scattering plays an important role to provide extra channels for heat transport, either due to the electron-phonon interaction or phonon-phonon interaction. Inelastic scattering even dominates the interfacial thermal transport when the phonon spectra of the two sides of interfaces are highly dissimilar [69].

Starting from different theoretical frames, several results suggests that inelastic scattering is possible to enhance the ITC [31, 70-72]. The inelastic scatterings provide extra channels for phonons to transport across the interface [70]. It also thermalize the phonons, which is possible to increase the population of modes with high transmission coefficients [71]. Another study reveals that the nonlinear coupling at the interface can enhance the ITC if the linear coupling is weak. On the contrast, if the linear coupling is strong, it will suppress the ITC, instead of increasing ITC [72].

## 4. Applications of ITC

### 4.1 Thermal rectification

An important consequence of inelastic scattering is the thermal rectification [24, 37], which plays the key role in the application of thermal diode [22] and phononics [23]. Thermal rectification is the effect that the thermal conductance differs between forward and backward flow of heat current. Interface scatterings can be used to realize such effects [24]. The significance of such effects is characterized by the thermal rectification ratio

$$R = (J_+ - J_-)/J_-$$

where $J_+$ ($J_-$) is the forward and backward heat flux under the same small temperature bias. Numerical simulations suggests that thermal rectification can be realized in many interfacial system, such as graphene-silicon junction [55], interfaces between suspended and encased graphene [50], silicon-amorphous polyethylene interface [73], metal-insulator interface via electron-phonon interaction [74], carbon isotope doping induced interface [56], two-dimension Ising lattice [75], one-dimensional anharmonic atomic chains [37, 76]. The rectification ratio can be up to 40-50% in material simulations [50, 73]. In one-dimensional model, it has been shown that the rectification results from the biased transmission properties of high-frequency phonons. Hence, thermal rectifier can be constructed by filtering the high frequency phonons in one direction.

### 4.2 Thermal interface materials

In the nano-electronic devices, heat is unavoidably produced. Fast dissipation of heat is one of the major issues that affect the performance and its maximum sustainable power. In the heat dissipation process, the interfacial thermal resistance is a challenge. The research of interfacial thermal materials aims to find materials that can be inserted into the devices and heat sink, in order to suppress the interfacial thermal resistance and enhance heat dissipation.

Some low-dimensional materials, such as graphene, show promising potentials in the applications of heat removal.

## 5. Summary and Outlook

In the continuous scaling down of nano-device, the management of interfacial thermal properties becomes more important to dates. The goal is that the heat flux across the interface can be effectively controlled, and thus the heat generated in the nano-devices is well managed. To achieve that, a deep and clear understanding of the transport mechanism at the interface is on demand. The theories of interfacial thermal transport have been continuously progressing, from the acoustic/diffusive mismatch model to the atomic-level modern simulation tools. The advances of theory and simulation power allow deeper understanding of the underlying interfacial transport mechanism, together with the merge of new research areas such as phononics [23].

Despite the numerous progresses, the interfacial thermal transport properties are still not fully understood, especially in the inelastic scattering regime. Molecular dynamics simulation can consider the anharmonic impacts; however, it is difficult to integrate the quantum effects. The inelastic quantum scattering at the interface is a challenge to capture. The quantum theories, such as NEGF, only allow perturbation treatment for the inelastic scattering, which will breakdown in the strong nonlinear regime. The quantum master equation formalism allows exact treatment of inelastic scattering [77, 78], but the computational complexity limits its application to small system with few degrees of freedom. Hence a comprehensive theory which fully describes the transport mechanism at the interface is still lacking. Such theory may enrich the physics of interfacial thermal transport properties and probably reveal new transport mechanisms.

Another important aspect of interfacial thermal transport is the metal-insulator interface, where the electron-phonon interaction plays an important role. In metal-insulator interface, besides the channel of lattice vibration, the thermal energy of metal can be transmitted through electron-phonon interaction. The electron-phonon interaction can take place in two different ways. The first one is electrons in metal first interact with the phonons in metal and then transmit the energy though lattice vibration at the interface. The other way is the direct interaction between the electrons in metal and the phonons in insulator. The electron-phonon interfacial conductance has been found substantially important[18, 29, 74, 79, 80]. The heat flow can be mediated through the surface state of the electrons [81]. However, simulation of electron-phonon interfacial thermal conductance is challenging due to its nonlinear nature and the strong quantum effects of electrons. Two temperatures Boltzmann transport equation approach has been developed for such problem based on the diffusive transport assumption, in combination of molecular dynamics simulation [82, 83]. In the weak electron-phonon interaction regime, NEGF provides a way to handle the transport problem in the junction setup [41]. However, a complete quantum mechanical description of electron-phonon interfacial problem is lacking, and also promising.

**ACKNOWLEDGMENT**

This work was supported in part by a grant from the Science and Engineering Research Council (152-70-00017). The authors gratefully acknowledge the financial support from the Agency for Science, Technology and Research (A*STAR), Singapore.

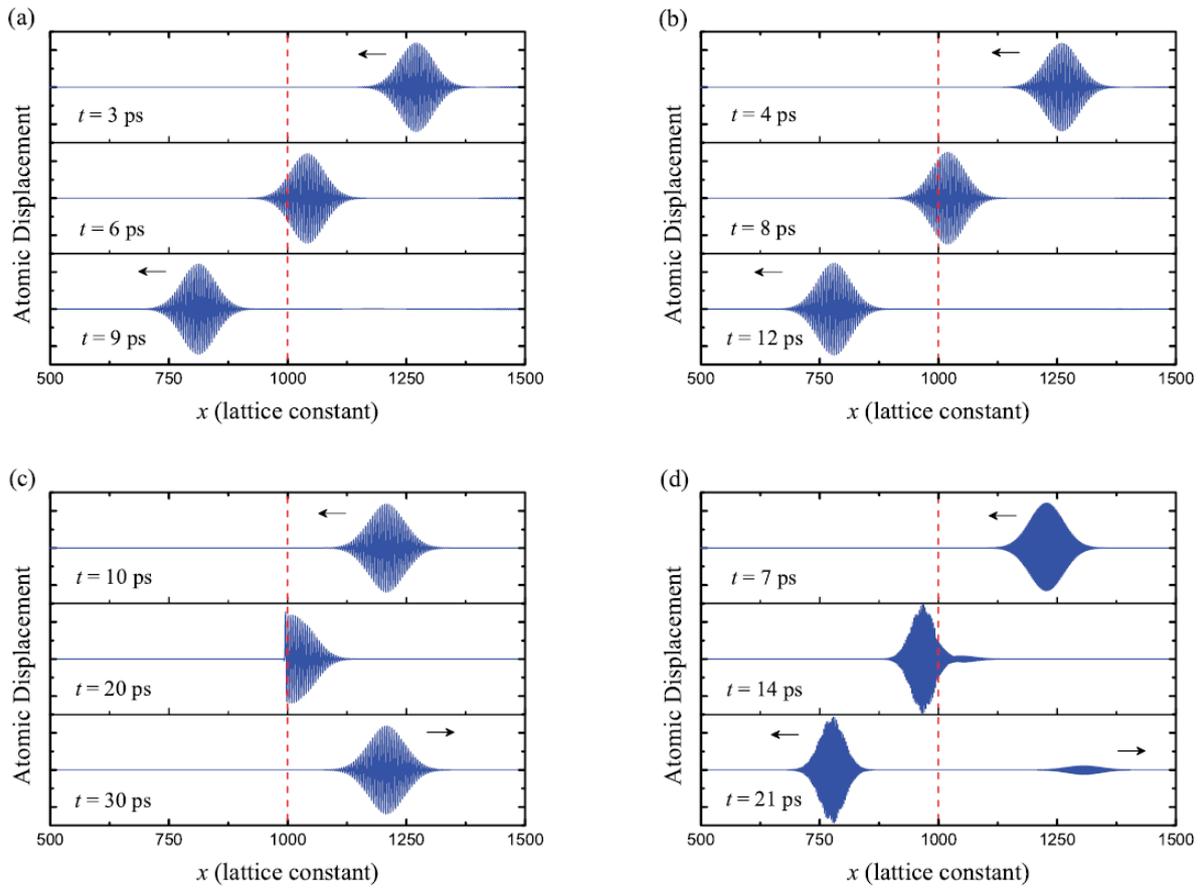

Figure 1. Snapshots of wave packets crossing the interface, red dashed lines denote the interface and the arrows indicate the travelling direction. (a) LA, frequency 16.8 THz. (b) TA, frequency 12.0 THz. (c) ZA, frequency 2.5 THz. (d) ZA, frequency 9.0 THz. Reprinted with permission from Ref [50], copyright (2014) by the American Institute of Physics.

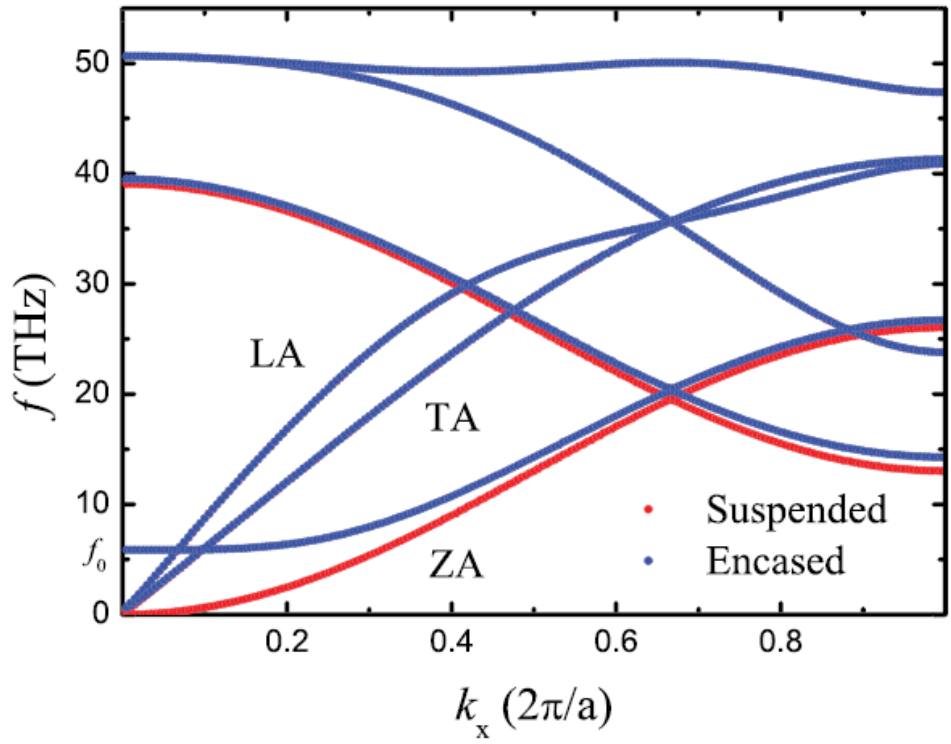

Figure 2. Phonon dispersions of suspended (red dots) and encased (blue dots) graphene. Reprinted with permission from Ref [50], copyright (2014) by the American Institute of Physics.

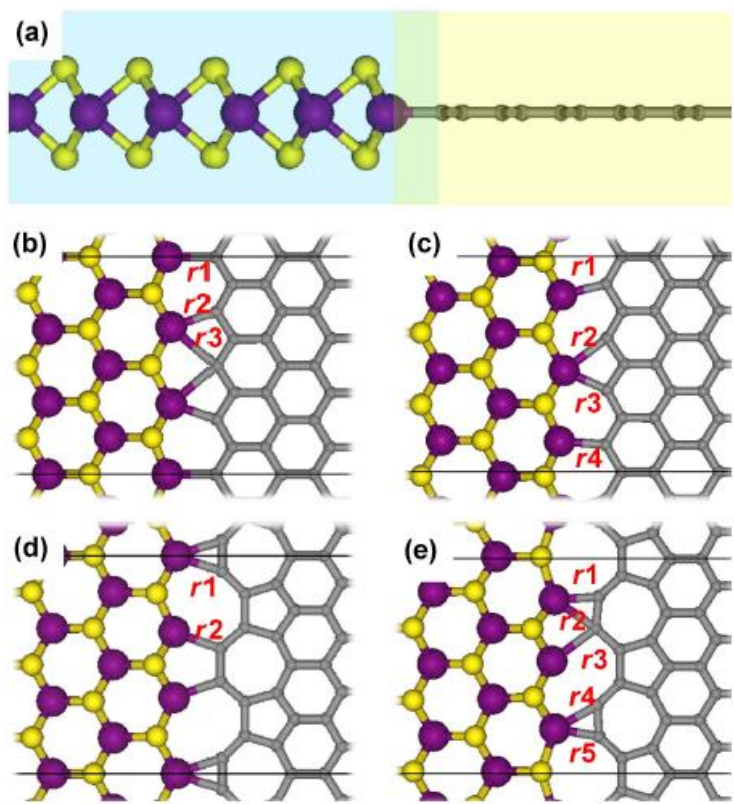

Figure 3. Interfaces of in-plane graphene-MoS2 heterostructures. Mo, S, and C atoms are shown in purple, yellow, and gray, respectively. Four possible interfacial configurations exist at the interface, as shown in (b)-(e). Reprinted with permission from Ref. [17], copyright (2017) by the Springer.

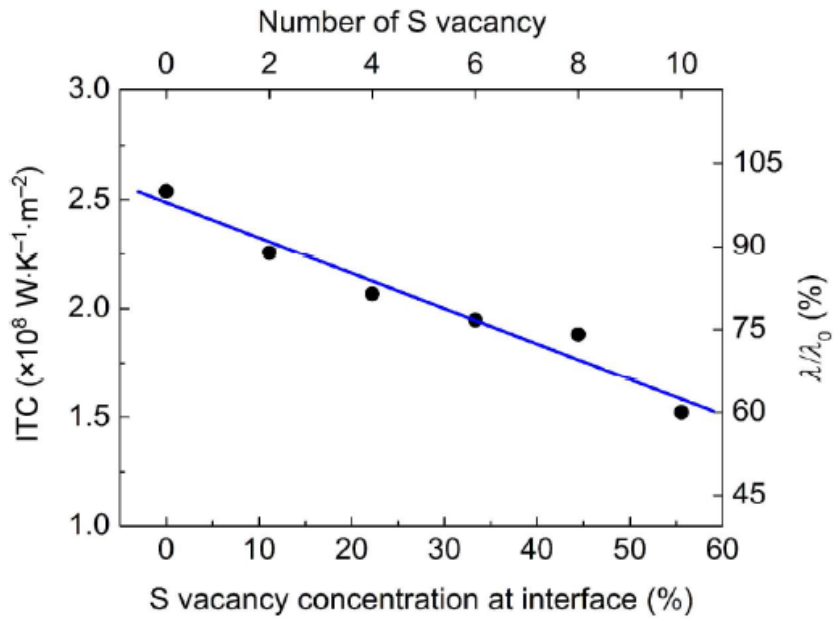

Figure 4. Interfacial thermal conductance as a function of S vacancy concentration at the interface. Reprinted with permission from Ref. [47] copyright (2016) by Springer.

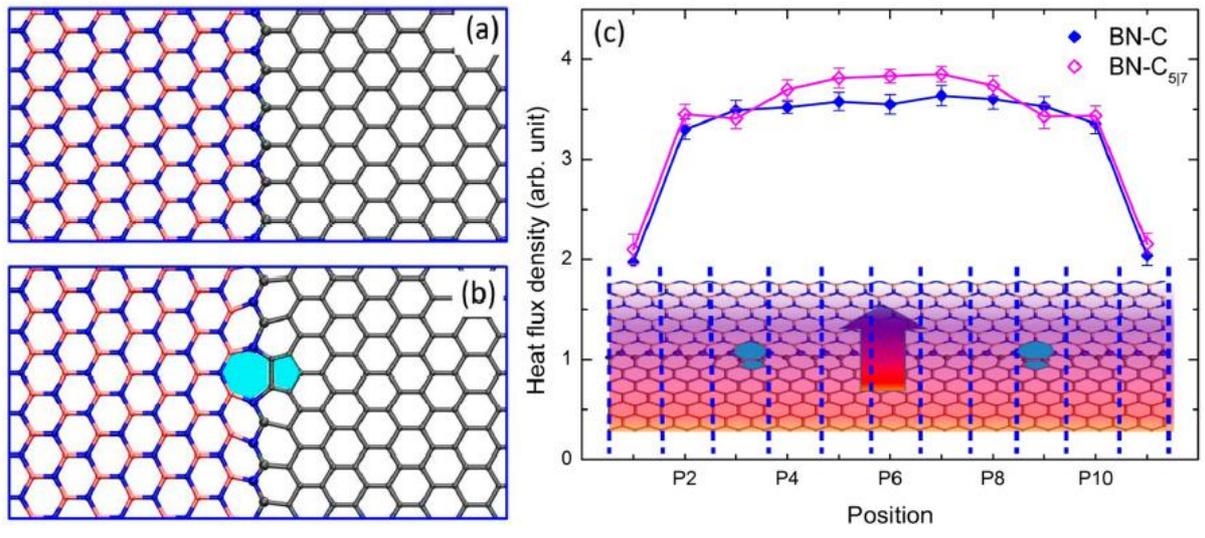

Figure 5. Heterostructure with different interfaces between graphene and h-BN. (a) N−C connected interface, BN−C; (b) N−C connected interface with 5|7 topological defects, BN−C5|7. B, N, C atoms are shown in purple, blue, and gray color, respectively. (c) Cross-sectional heat flux density distribution in BN−C and BN−C5|7 heterostructures. Reprinted with permission from Ref. [84], copyright (2016) by the American Chemical Society.

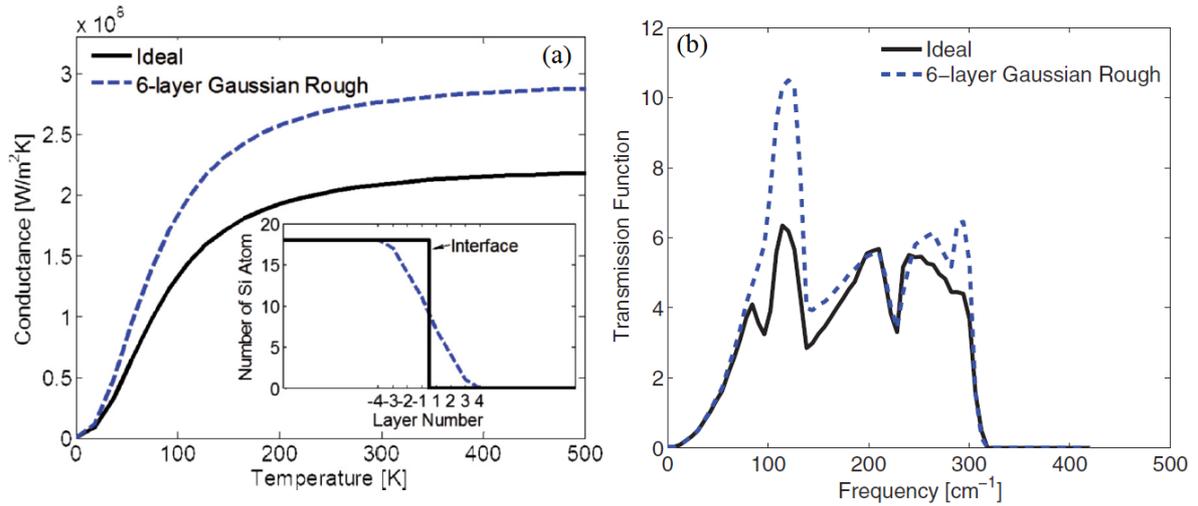

Figure 6. (a) Thermal conductance as a function of temperature for an ideal Si/Ge interface (solid black line) and for a rough Si/Ge interface with a Gaussian distribution (dashed blue lines). (b) Total transmission function for an ideal Si/Ge interface and for a rough Si/Ge interface. Reprinted with permission from Ref. [54], copyright (2012) by the American Physical Society.

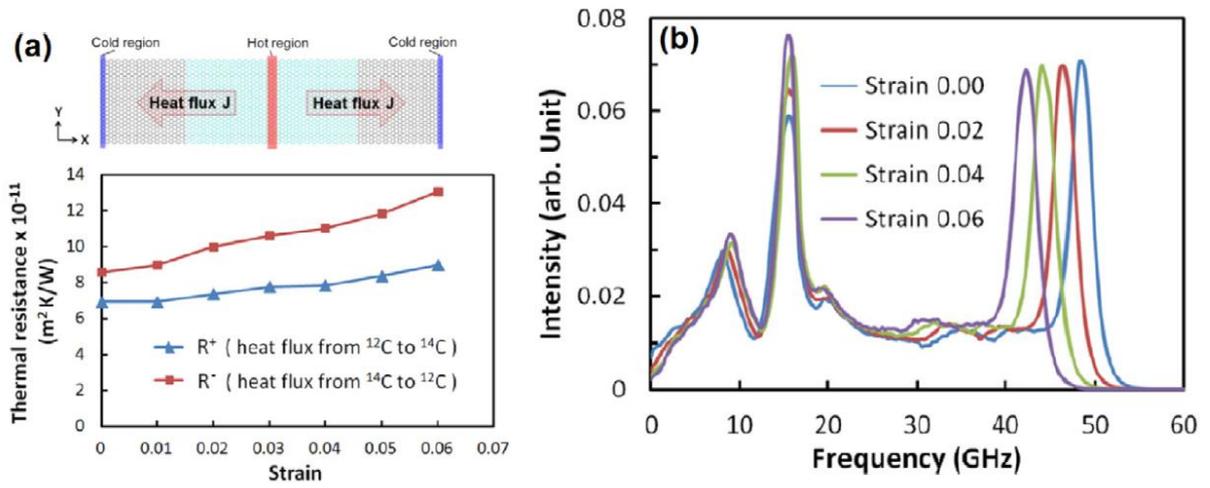

Figure 7. (a) The effect of tensile strain on the interfacial thermal resistance. The interface between pristine and isotope doped graphene is also shown here. (b) Phonon spectra of atoms near the interface under different tensile strains. Reprinted with permission from Ref [56], copyright (2012) by the American Institute of Physics.

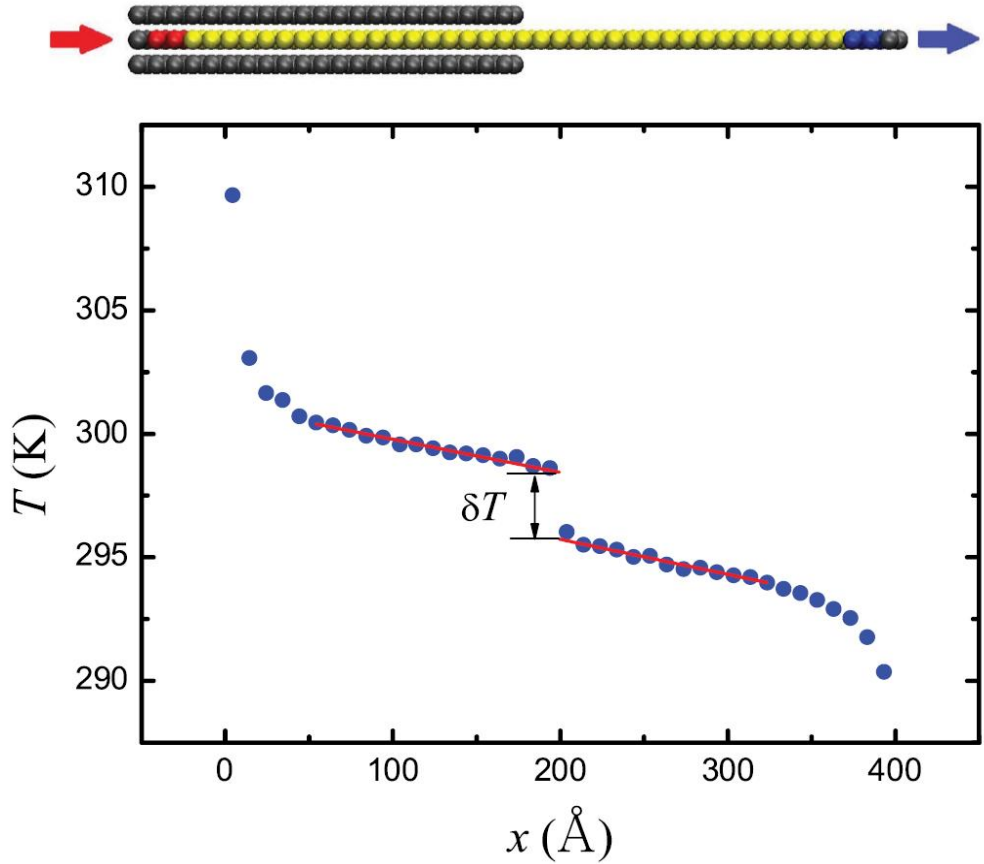

Figure 8. Side view of the simulating model, and a typical temperature profile at steady state. Here the black colored atoms are fixed, red and blue colored ones contact with heat source and sink in non-equilibrium MD simulations, respectively. Reprinted with permission from Ref [50], copyright (2014) by the American Institute of Physics.